\documentstyle[12pt]{article}
\input epsf
\begin{document}

\begin{center}
\large{{\bf Comments on ``Dark matter: A phenomenological existence proof''}}
\end{center}
                                                                                
\begin{center}
R. G. Vishwakarma,\footnote{E-mail: rvishwa@mate.reduaz.mx, rvishwak@ictp.it}\\

\medskip
\small{Unidad Acad$\acute{e}$mica de Matem$\acute{a}$ticas\\
 Universidad Aut$\acute{o}$noma de Zacatecas\\
 C.P. 98068, Zacatecas, ZAC.\\
 Mexico}\\
\end{center}

\medskip
\begin{abstract}

\noindent
A recent paper by Ahluwalia-Khalilova \cite{ahluwalia} is examined where
he claims that the standard FRW cosmology with a $\Lambda$ predicts existence
of dark matter without invoking the data on galactic rotation curves and
gravitational lensing. However, we find that his claims are not correct.
He has already assumed (without realizing) in the very outset what he wants 
to prove.

\end{abstract}

\bigskip
 
\noindent
Recently Ahluwalia-Khalilova has claimed \cite{ahluwalia} that the flat 
Friedmann-Robertson-Walker (FRW) cosmology with a cosmological constant 
$\Lambda$ predicts the existence of dark 
matter $\Omega_{\rm dm0}\approx 0.22\pm0.08$, without invoking the data on 
galactic rotation curves or gravitational lensing
(where, as usual,  $\Omega_{\rm i}$ is the energy density of the constituent
$\rho_{\rm i}$ in units of the critical density $3H^2/8\pi G$ and the 
subscript `0' denotes the value of the quantity at the present epoch). 
However, this is not a 
prediction, but just a reflection of our assumption once
we consider the existence of a $\Lambda$ ($\Omega_{\Lambda0}\approx 0.73$) 
in a flat universe together with the insufficient visible matter 
$\Omega_{\rm baryon}\approx 0.05$ to close the universe. However, 
Ahluwalia-Khalilova also claims that he has not used any input from $\Lambda$
(except for working in units of $1/\sqrt{\Lambda}$  for convenience). We
find that this is not correct. The constraint on $\Omega_{\rm m0}$ found by 
him has not been derived without using $\Omega_{\Lambda0}\approx 0.73$, as
will be explained in the following.

In the standard FRW cosmology, the present universe is described by 
the non-relativistic matter (of energy density $\rho_{\rm m}$) and a constant 
$\Lambda$. The dynamics of the evolution is given by the  the Einstein field 
equations 
\begin{equation}
H^2\equiv \left(\frac{\dot{a}}{a}\right)^2 = \frac{8\pi G}{3}\rho_{\rm m} + 
      \frac{\Lambda}{3}-\frac{k}{a^2},  
      \label{eq:1} 
\end{equation}
\begin{equation}
\dot{\rho}_{\rm m} + 3\rho_{\rm m}\left(\frac{\dot{a}}{a}\right)=0,  
      \label{eq:2} 
\end{equation}
which can be integrated easily to give the age of the universe 
\cite{vishwakarma}:
\begin{equation}
t_0=\frac{1}{H_0}\int_0^\infty \frac{dz}{\sqrt{[\Omega_{\rm m0}(1+z)^5-
(\Omega_{\rm m0}+\Omega_{\Lambda0}-1)(1+z)^4+\Omega_{\Lambda0}(1+z)^2]}}.
\label{eq:age}
\end{equation}
 It should be noted that at its face value the age $t_0$ appearing in 
equation (\ref{eq:age}) is, de facto, 
the dynamical age of the universe (the time it took the universe to evolve 
from the big bang to the present epoch) which is clearly a function of the 
parameters $H_0$, $\Omega_{\rm m0}$ and $\Omega_{\Lambda0}$. As given by 
equation (\ref{eq:age}), it {\it does not} represent the age
of the constituents of the universe, like globular clusters or stars.
The age of the universe must be more than the age of its constituents 
(the structures did not start forming at the big bang).
Of course one can always check if the age of the universe is consistent with 
the age of its constituents. However, the age of even the oldest objects in 
the universe (for example, the globular clusters) should provide only
the lower limit.   

Further, a given set of the parameters  $H_0$, $\Omega_{\rm m0}$ and 
$\Omega_{\Lambda0}$, gives a unique value of $t_0$ from equation 
(\ref{eq:age}), but the reverse case is very degenerate in solutions, even in 
the above mentioned limiting case. As there are 4 unknowns  
in equation (\ref{eq:age}), equating $t_0$ to the age of a particular 
object  can yield
many sets of $H_0$, $\Omega_{\rm m0}$, $\Omega_{\Lambda0}$ as solutions.
For example, the values $\Omega_{\rm m0}=0.271\pm0.028=1-\Omega_{\Lambda0}$ 
(inferred by Astier 
et al. \cite{astier} which is considered by Ahluwalia-Khalilova in his paper)
and $H_0=72\pm8$ km s$^{-1}$ Mpc$^{-1}$ (from the HST key project \cite{hst}),
give a $t_0=13.48\pm1.88$ Gyr which is consistent with the age of the globular
clusters $t_{\rm GC}=13.5\pm1.5$ Gyr used by Ahluwalia-Khalilova. However,
another model $\Omega_{\rm m0}=\Omega_{\Lambda0}=0$ with the same value
of $H_0$ also gives a $t_0=13.58\pm1.51$ Gyr which is consistent with the value
of $t_{\rm GC}$ mentioned above. Further, one can find many other solutions 
permitted by the uncertainties in the observed values of the parameters.

Even after restricting ourselves to the flat models $\Omega_{\rm m}
+\Omega_\Lambda=1$ (as has been done by
Ahluwalia-Khalilova), the solutions will be degenerate in 2 parameters. 
Thus, just by setting a value to $t_0$, one cannot solve for $\Omega_{\rm m0}$
(or $\Omega_{\Lambda0}$) uniquely unless one also sets some value to $H_0$ 
(possibly from some other observation). 
Thus all one can do is one can check whether
a particular value of $t_0$ (say, the age of the globular clusters) is 
consistent with a model with a certain $\Omega_{\rm m0}$ together with a 
certain $H_0$. However, this cannot be regarded as a prediction or a proof
of existence of that $\Omega_{\rm m0}$. It is just a consistency check, as
mentioned earlier. Ahluwalia-Khalilova has also done a similar consistency
check, as will be explained in the following, though he does not realize it.

Ahluwalia-Khalilova has considered, in his paper \cite{ahluwalia}, the
analytical solution of equations (\ref{eq:1}) and  (\ref{eq:2})
available in the literature, viz.
\begin{equation}  
a^3(t)=\frac{8\pi G \rho_{\rm m0}a_0^3}{\Lambda}\sinh^2
\left(\frac{\sqrt{3\Lambda}}{2}t\right),\label{eq:sfactor} 
\end{equation}
for the case $k=0$. By the use of the solution of the conservation equation 
(\ref{eq:2}) (i.e., $\rho_{\rm m}a^3=$ constant 
$=\rho_{\rm m0}a_0^3$, say), equation (\ref{eq:sfactor}) can be put in the form
\begin{equation}  
t=\frac{2}{\sqrt{3\Lambda}} \sinh^{-1}\sqrt{\frac{\Omega_\Lambda(t)}{\Omega_{\rm m}(t)}},
\end{equation}
giving
\begin{equation}  
t_0=\frac{2}{\sqrt{3\Lambda}} \sinh^{-1}\sqrt{\frac{1}{\Omega_{\rm m0}}-1}.\label{eq:age1}
\end{equation}
Now Ahluwalia-Khalilova claims that in order to evaluate $\Omega_{\rm m0}$ 
from equation (\ref{eq:age1}), he uses the only input from the observations
as $t_0=t_{\rm GC}$, and does not take any input from $\Lambda$ except for 
working in units of $1/\sqrt{\Lambda}$ ($\approx$9 Gyr) for convenience. 
However this 
is not correct. He does use the observed value of $\Lambda$ in equation 
(\ref{eq:age1}) without realizing it, as we shall see shortly. 

If one wants to work in units of $1/\sqrt{\Lambda}$ for the sake of 
convenience, the proper way is to consider $1/\sqrt{\Lambda}$ in units of 
$10^9$ years. So, let us consider $1/\sqrt{\Lambda}=\lambda\times 10^9$ 
years, where $\lambda$ is
the dimensionless value of  $1/\sqrt{\Lambda}$ in units of $10^9$ years.
Equation (\ref{eq:age1}) then gives
\begin{equation}  
\Omega_{\rm m0}=\left[ 1+\sinh^2\left\{\frac{\sqrt{3}}{2\lambda}(13.5\pm1.5)\right\}\right]^{-1}.\label{eq:omega}
\end{equation}
Now, can one evaluates the r.h.s. of equation (\ref{eq:omega}) without putting
a numerical value of $\lambda$ and can get $\Omega_{\rm m0}=0.27\pm0.08$, as
obtained by Ahluwalia-Khalilova? The answer is a flat no. So, a value of
$\Lambda$ has been assumed by him a priori!

It should be worthwhile here to comment on the observational status of
$\Lambda$. It should be noted that $\Lambda$ (or any other variant of dark 
energy) is not observed directly like the visible matter density
($\rho_{\rm baryon}$). Rather it is inferred from the observations by comparing
the distance (luminosity distance or angular diameter distance) of an object
predicted by a cosmological model to the observed distance given by the 
observations of supernovae of type Ia, cosmic microwave background or 
gravitational lensing. One can also infer bounds on it by comparing the dynamical age
of the universe to the age of the oldest objects in it. These methods
estimate $\Omega_{\Lambda0}$, from which $\Lambda$ is deduced by using the
observed value of the Hubble parameter $H_0$. Thus, in quoting a value for
$\Lambda$ from the observations, there is always an associated $H_0$ tacitly
assumed. This fact is
reflected when one writes equation (\ref{eq:age1}) in the following 
alternative form by using the relations 
$\Lambda/3H_0^2=\Omega_{\Lambda0}=1-\Omega_{\rm m0}$: 
\begin{equation}  
t_0=\frac{2}{3H_0\sqrt{1-\Omega_{\rm m0}}} \sinh^{-1}\sqrt{\frac{1}{\Omega_{\rm m0}}-1},\label{eq:age2}
\end{equation}
which gives the same answer as equation (\ref{eq:age}) for the case $k=0$. 

Hence it is clear that by assuming a $\Lambda$ (from $1/\sqrt{\Lambda}\approx
9$ Gyr), Ahluwalia-Khalilova has already assumed the existence of 
an $\Omega_{\Lambda0}\approx$ 0.73 (for $H_0\approx72$ km s$^{-1}$ Mpc$^{-1}$)
and hence an $\Omega_{\rm m0}=1-\Omega_{\Lambda0}$ ($\approx$ 0.27). It is already known that these values give the
correct age of the universe. Thus the exercise done by 
him is simply a consistency check of the standard cosmology
and does not provide any existence proof for $\Omega_{\rm m0}$ and hence
for dark matter (in view of the insufficient 
$\Omega_{\rm baryon}\approx 0.05$ to close the universe).  

It should also be noted that the value $\Lambda/(8\pi G)
\approx 4 \times 10^{-47}~\mbox{GeV}^4$ does not seem to give 
$\tau_\Lambda \equiv1/\sqrt{\Lambda}\approx 9$ Gyr, as used by 
Ahluwalia-Khalilova. Rather it should be replaced by 
$\Lambda/(8\pi G)\equiv \rho_{\rm v}\approx 3 \times 10^{-47}~\mbox{GeV}^4$
(in units of $c=1$, $\hbar=1$).

\end{document}